\documentclass[prl,reprint,showpacs]{revtex4-1}
\usepackage[pdftex]{graphicx}  
\usepackage{dcolumn}   
\usepackage{bm}        
\usepackage{amssymb}   
\usepackage{amsmath}   
\usepackage{hyperref}
\usepackage[latin1]{inputenc}
\usepackage{tikz}
\usetikzlibrary{shapes,arrows}

\begin{document}

\title{The role of electron localization in density functionals} 
\date{\today}
\author{M.\ J.\ P.\ Hodgson}
\author{J.\ D.\ Ramsden}
\author{T.\ R.\ Durrant}
\author{R.\ W.\ Godby}
\affiliation{Department of Physics, University of York and European Theoretical Spectroscopy Facility, Heslington, York YO10 5DD, United Kingdom} 

\begin{abstract}
We introduce a new functional for simulating ground-state and time-dependent electronic systems within density-functional theory.
The functional combines an expression for the exact Kohn-Sham (KS) potential in the limit of complete electron localization with a measure of the actual localization. We find accurate self-consistent charge densities, even for systems where the exact exchange-correlation potential exhibits non-local dependence on the density, such as potential steps. 
We compare our results to the exact KS potential for each system. The self-interaction correction is accurately described, avoiding the need for orbital-dependent potentials.
\end{abstract}


\maketitle

Density-functional theory (DFT) \cite{PhysRev.136.B864} is the most widely used tool for the simulation of many-electron systems in numerous fields of physics, chemistry and materials science. Its success hinges on approximations \cite{PhysRev.140.A1133,PhysRevLett.77.3865} to the exchange-correlation (xc) part of the Kohn-Sham (KS) functional, which perform well across a range of ground-state systems. However, these approximations become much less secure in the presence of strong correlation \cite{PhysRevLett.102.026101,PhysRevB.23.5048} and/or current flow \cite{PhysRevB.73.121403, DFcalc,PhysRevLett.109.036402, PhysRevLett.109.266404, PhysRevB.88.241102}. Particular attention has been given to improving the time-dependent xc potential, used within time-dependent DFT (TDDFT) \cite{PhysRevLett.52.997}, where the use of adiabatic functionals of the electron density ignores the role of currents and memory effects.

In this Letter we demonstrate that electron localization, driven by the Coulomb interaction and the Pauli principle, can form a powerful ingredient in approximations for the KS potential. The electron localization function (ELF), $L(x)$, as in Ref.~\onlinecite{dobson1991interpretation}, provides a useful indicator of localization: $L = 1$ is complete localization, i.e.\ the chance of finding one electron in the vicinity of another is zero. $L(x)$ ranges from $0$ to $1$, and a homogeneous electron gas (HEG) has $L = 0.5$.

Our starting point is the KS potential of Refs.~\onlinecite{smith1979density} and \onlinecite{hessler2002correlation}, originally derived for a system of two spinful electrons in their spin-zero ground state. We observe that the logic applies exactly to any one-electron system, and, indeed, in a general system, to all regions of space where the electron density is dominated by any one Kohn-Sham orbital 
\footnote{In such situations, symmetry may cause the exact KS orbitals to extend throughout the system. As the localized limit  is approached these orbitals become degenerate and can be rotated within  the Hilbert space onto effectively localized orbitals while remaining eigenfunctions of the Hamiltonian.}. For such a region the KS equations may be approximated as \footnote{We use Hartree atomic units.} $\left ( -\tfrac{1}{2} \triangledown^2 + V_\mathrm{KS} \right )  \sqrt{n} = \varepsilon_k \sqrt{n}$ for the dominant orbital $\phi_k$, where $n \approx |\phi_k|^2$ in the region, yielding the ground-state KS potential, which we term the \textit{single orbital approximation} (SOA),
\begin{equation}
V^{\mathrm{SOA}}_{\mathrm{KS}}(\textbf{r}) = \frac{\triangledown^2 n(\textbf{r})}{4n(\textbf{r})} -\frac{[\triangledown n(\textbf{r})]^2}{8n(\textbf{r})^2}.
\label{SOA}
\end{equation}
(Here the zero of energy in the KS system is at $\varepsilon_k $.)

We begin by considering Eq.~\ref{SOA} as an approximation to the universal KS functional. We find that the SOA not only works well for the strongly localized orbital regions, but also accounts for non-local features and corrects self-interaction in the KS potential in regions of low localization. We compare the SOA to the exact KS potential for a variety of ground-state and time-dependent systems that exhibit non-local behavior in the xc potential. We then extend our approach by combining the SOA with a potential suited to delocalized systems, in proportions depending on the strength of localization: the mixed localization potential (MLP), introduced in Eq.~\ref{MLP+ref} below, which is amenable to self-consistent use in DFT calculations.

Our four test systems consist of either two or three spinless electrons in one dimension of varying degrees of electron localization. Spinless electrons maximize the richness of correlation for a given computational effort, with each electron occupying a distinct KS orbital. 
The electrons interact through the appropriately softened Coulomb repulsion $1/(|x-x'|+1)$.
The calculations are performed using our iDEA code \cite{PhysRevB.88.241102}, which determines the exact KS potential from the fully correlated ground-state or time-dependent many-electron wavefunction. 

\textit{System 1 (double well)}.---We begin by studying a ground-state system where the electrons are highly localized: two spinless electrons subject to an external potential consisting of two identical spatially-separated wells together with a potential step between them (Fig.~\ref{DoubleWell}(a)) \footnote{See Supplemental Material at [URL will be inserted by publisher] for specific parameters of our test systems.}.
In the absence of interaction, the electrons would simply occupy the two lowest single-particle states, which are both located in the left-hand well. The Coulomb repulsion, however, localizes the electrons, with one electron per well. To reflect this (as originally noted by Almbladh and von Barth \cite{almbladh1985density}), the xc part of the KS potential must incorporate a second spatial step between the wells, ensuring one electron per well through the ordered filling of the KS orbitals. 

We calculate the exact KS potential for this system and compare to that given by the SOA (Eq.~\ref{SOA}), evaluated for the exact density (Fig.~\ref{DoubleWell}(b)). 
The anticipated new step in the exact KS potential (arrow in Fig.~\ref{DoubleWell}(b)) appears, located at the density minimum: the ``interface'' between the two electrons where the localization is lowest.  The SOA replicates this xc step quite well \footnote{Tests indicate that the step in the exact and SOA potentials always forms at the density minimum}, despite being at its least secure in this region of low localization. The SOA's ability to form this potential step is particularly impressive in this region of  very low electron density, showing proper sensitivity of Eq.~\ref{SOA}  to tiny variations in the density (inset in Fig.~\ref{DoubleWell}(b)).
The self-interaction correction, which is the dominant new feature in the exact $V_\mathrm{KS}$ within the wells, is very well reproduced by the SOA, owing to its exactness in the limit of complete localization. 

\begin{figure}[htbp]
  \centering
  \includegraphics[width=1.0\linewidth]{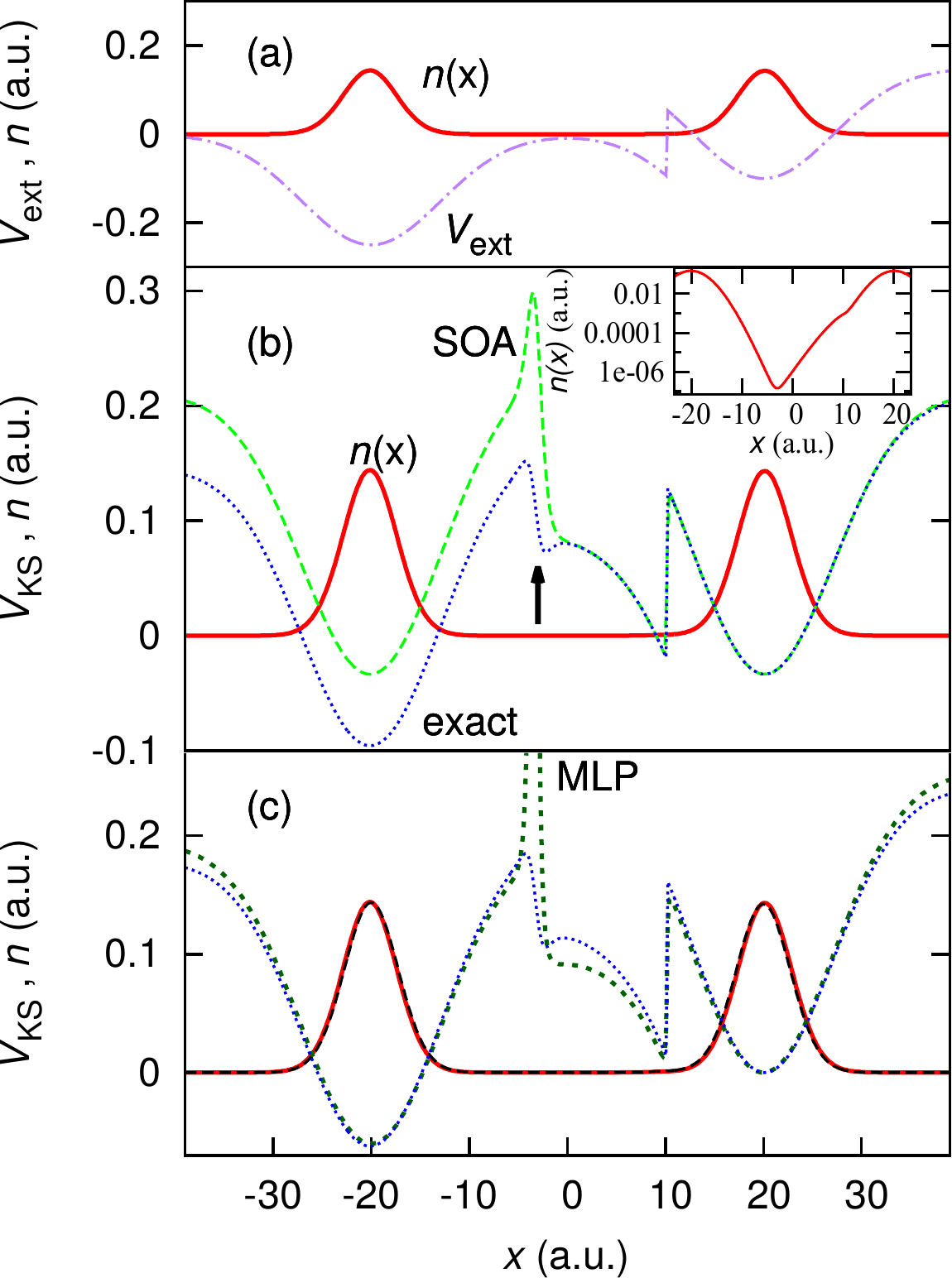}
\caption{(color online) \textit{System 1 (double well)}. (a) The exact many-body electron density (red solid line for all three plots) with the external potential (dotted-dashed purple). A potential step gives the right-hand well a lower ionization energy, while the Coulomb repulsion ensures one electron per well. (b) The exact KS potential (blue dotted, also in (c)); an xc potential step (arrow) forms at the minimum in the electron density. The SOA potential (green dashed) replicates the position and magnitude of the step well. Inset: detail of the density in the neighborhood of the two steps. The delicate features that give rise to the steps in $V_\mathrm{KS}$ are visible at $x=-3$ and $x=10$. (c) The \textit{self-consistent} MLP potential, for $f=0.6$, (short-dashed dark green) with the corresponding electron density (dashed black), which matches the exact density very well.}
\label{DoubleWell}
\end{figure}

A useful approximate functional must give accurate densities when applied self-consistently without prior knowledge of the exact density. The SOA, as it is ``unanchored'' to the external potential, is not suited to this, but by mixing the SOA with a suitable reference potential we can better approximate the KS potential: 
\begin{equation}
V^{\mathrm{MLP}}_{\mathrm{KS}} = f V^{\mathrm{SOA}}_{\mathrm{KS}} + f' V^{\mathrm{ref}}_{\mathrm{KS}},
\label{MLP+ref}
\end{equation}
where $f(x)$ and $f'(x)$ depend on the localization. We term this the \textit{mixed localization potential} (MLP).

For complete localization ($L=1$),  $f=1$ and $f'=0$. Similarly for the HEG ($L=0.5$)  $V^{\mathrm{ref}}_{\mathrm{KS}}=V^{\mathrm{LDA}}_{\mathrm{KS}}$, and hence $f=0$ and $f'=1$. The precise mapping between $L$ and $f$ must be that appropriate to the reference potential used. We can also infer, from the sum rule of the xc hole \cite{PhysRevB.13.4274}, that $f+f' \approx 1$.

For a highly localized system, such as System 1, the failure of the LDA to correct the self-interaction would make its use as a reference potential damaging in the regions of high density, and in the regions of low density any LDA-like potential will have small Hartree and exchange-correlation (Hxc) terms. Thus $V_\mathrm{ext}$ is a superior choice of reference potential \footnote{In this case, we can approximate Eq.~\ref{MLP+ref} as
$V^{\mathrm{MLP}}_{\mathrm{KS}} = f V^{\mathrm{SOA}}_{\mathrm{KS}}+ (1-f) V_{\mathrm{ext}}
= V_{\mathrm{ext}}+f V^{\mathrm{SOA}}_{\mathrm{Hxc}}$, 
showing that the role of $f$ is to adjust the SOA Hxc potential.}.

We use the MLP to solve, \textit{self-consistently}, for two electrons in the double well system, using the standard potential-mixing iterative procedure of DFT. We choose the simplest approximation for $f(x)$, a constant throughout space. For multiple spinless electrons we must have some delocalization, especially at the ``interface'' between the electrons, implying that $f<1$. We have investigated a range of values of $f$, and find the density to be accurate when $f$ is in the range $0.6-0.9$ (Figure~\ref{DoubleWell}(c)). (Note that for this double-well system the KS potential is approximately the external potential plus a new step. Therefore, $V^{\mathrm{MLP}}_{\mathrm{KS}}$ is unaffected by assuming $f$ to be lower than $1$ for the regions of strong localization, as it is simply mixed with the external potential.) 

$f(x)=0.6$ (consistent with the partial delocalization in the interface region) reproduces the features of the exact $V_\mathrm{KS}$ with remarkable accuracy, including the alignment of the wells and the magnitude and position of both steps, correctly allowing one electron to occupy each well \footnote{For potentials with sufficiently high steps, the SOA correctly places \textit{both} electrons in the lower well, even if each KS electron begins in a different well at the start of self-consistency.}. (In addition to the step, $V^\mathrm{MLP}_{\mathrm{KS}}$ also exhibits a high, narrow peak, arising from $f$ being approximated as a constant, but this has little effect on the electron density, owing to the density being very small in this region.)

\begin{figure}[htbp]
  \centering
  \includegraphics[width=1.0\linewidth]{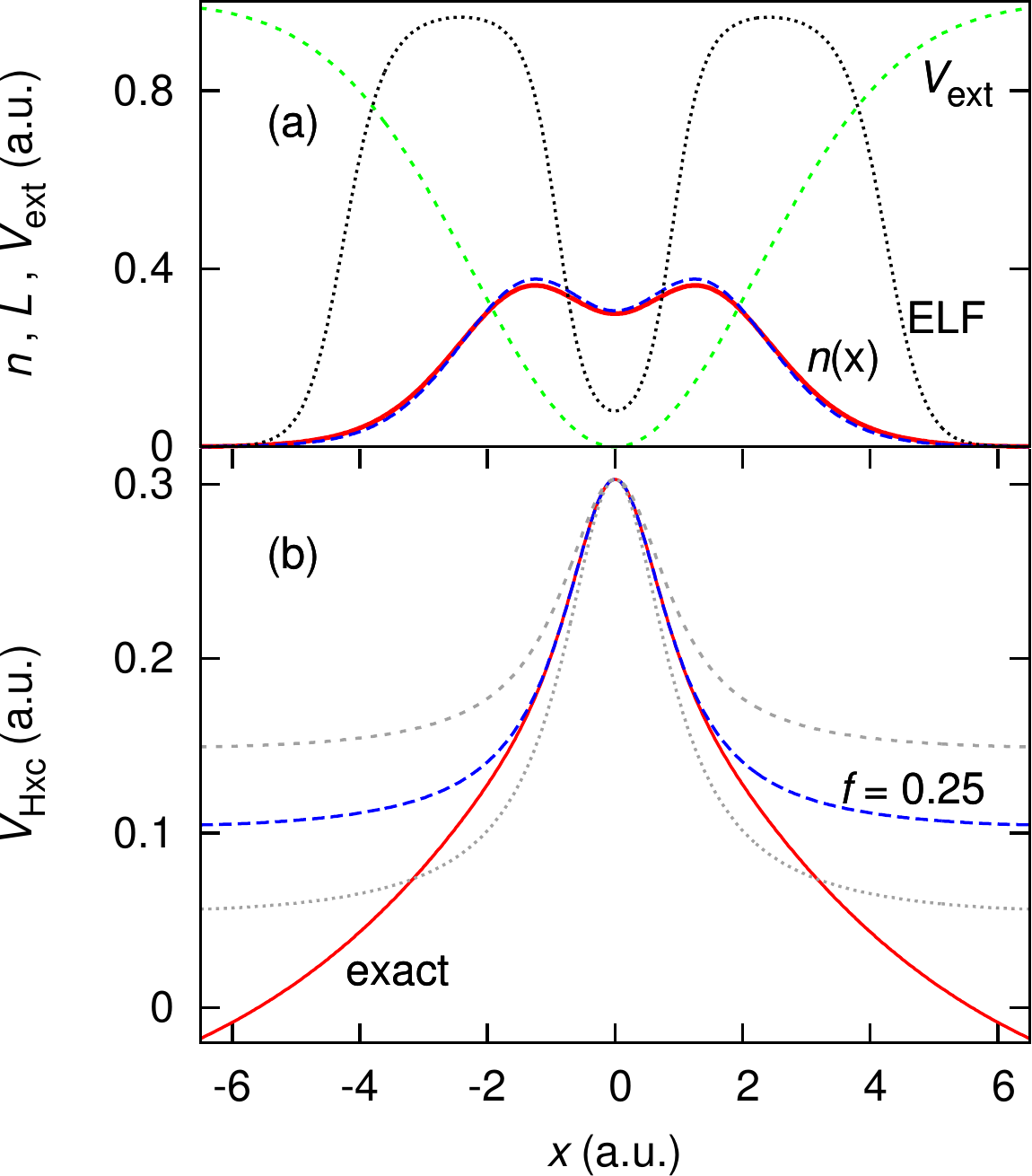}
\caption{(color online) \textit{System 2 (single well)}. (a) The many-body electron density (solid red), self-consistent MLP density (dashed blue), external potential (short-dashed green) and ELF (dotted black). Note the large dip in the localization of the electrons in the center of the well. (b) The exact Hxc potential (solid red) together with the Hxc potential from the self-consistent MLP, for $f=0.25$ (dashed blue; also $f=0.2$ and $0.3$ in gray). The MLP Hxc potential is in very good agreement with the exact in the center; the two potentials differ at the edges as a result of approximating $f$ as a constant.}
\label{Single Well}
\end{figure}

\textit{System 2 (single well)}.---Next we consider a system where the localization of the electrons drops \textit{significantly} below 1: two electrons confined to a single well. This will establish the performance of the MLP functional well outside the limit in which the underlying SOA is exact.
Figure~\ref{Single Well}(a) shows the ELF for this system together with the exact electron density and the MLP density, showing that the electrons are significantly delocalized in the central region.  In this region the exact potential has a ``bump'' which acts to push the density peaks apart, together with long-range Hxc fields on either side; the raw SOA contains both features, though their magnitude is overestimated in the delocalized region. 

In forming the MLP, we are guided by the ELF in the region of low localization and have explored a range of values for $f$ of $0.2-0.3$. For simplicity, we continue to use the external potential as our reference. The MLP yields high accuracy for the Hxc potential in this central region. 
The central bump in the potential is accurate across our range of $f$, and the corresponding densities are in good agreement with the exact. We find that $f=0.25$ reflects the delocalization in the central region (approximately the average of the ELF in this section), and hence yields a highly accurate $V^{\mathrm{MLP}}_{\mathrm{Hxc}}$ in the central region (see Fig.~\ref{Single Well}(b)). 

For the edges of the system the approximate and exact potentials differ; this is attributable to the approximation of $f(x)$ as a constant tailored to the central delocalized region, causing $V^{\mathrm{MLP}}_{\mathrm{Hxc}}$ to subdue the long-range Hxc fields. To improve this result one would need to have a spatially varying $f$, dependent on $L$, which introduces more of the SOA into this region. Nevertheless, even with a constant $f$ and a simple $V^\mathrm{ref}_\mathrm{KS}$ the MLP performs remarkably well, far from its exact limit.

\begin{figure}[htbp]
  \centering
  \includegraphics[width=1.0\linewidth]{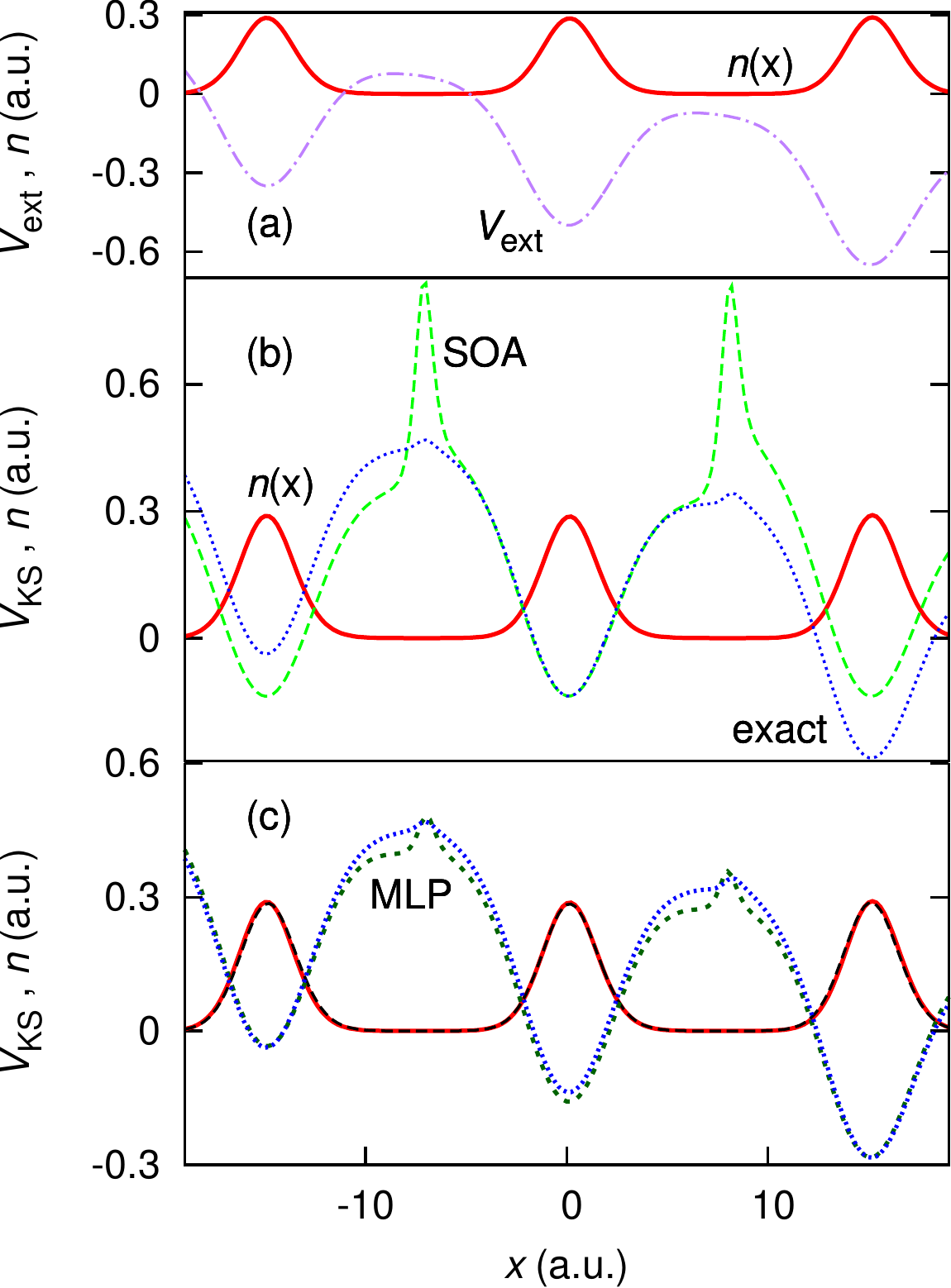}
\caption{(color online) \textit{System 3 (polarized three-atom chain)}. (a) The electron density for a chain of three atoms (one electron per well) in the many-body picture (solid red for each plot), with the external potential (dotted-dashed purple). (b) The exact KS potential (dotted blue, also in (c)); xc steps act to screen the external electric field. The SOA KS potential (dashed green) is also shown; this predicts complete screening of the electric field. (c) The self-consistent MLP density (dashed black) and KS potential (short-dashed dark green), mixing in the external potential with $f=0.17$.}
\label{3AtomChain}
\end{figure}

\textit{System 3 (polarized three-atom chain)}.---We study a chain of three atoms (one electron per well), with an applied electric field in the ground state (Fig.~\ref{3AtomChain}(a)). We chose this system to test the MLP primarily due to the charge imbalance predicted by the LDA, which fails to screen the field sufficiently \cite{van1999electric, PhysRevLett.74.4035}, since it lacks ultra-non-local density-dependence of the exact $V_{\mathrm{xc}}$ \cite{PhysRevLett.88.186401, van2003application}.

We find that the exact xc potential partly counteracts the applied external electric field, principally through two potential steps (Fig.~\ref{3AtomChain}(b)). The SOA predicts complete screening of the applied bias via similar steps, aligning the lowest KS eigenvalue within each potential well.

For this system, we would expect strong localization, except in the two crucial ``interface'' regions where the KS potential exhibits a step, and therefore use the external potential as reference as discussed for System 1. A range of values of $f(x)$ from $0.17$ to $0.77$, applied self-consistently in the MLP, yield accurate densities, with $f(x)=0.17$ yielding the most accurate screening in the MLP potential. (We note in passing that a position-dependent $f(x)$ that increases to 0.77 at the edges produces more accurate Hxc fields in the two outer wells, causing the density in these wells to become more accurately polarized.)

\begin{figure}[htbp]
  \centering
  \includegraphics[width=1.0\linewidth]{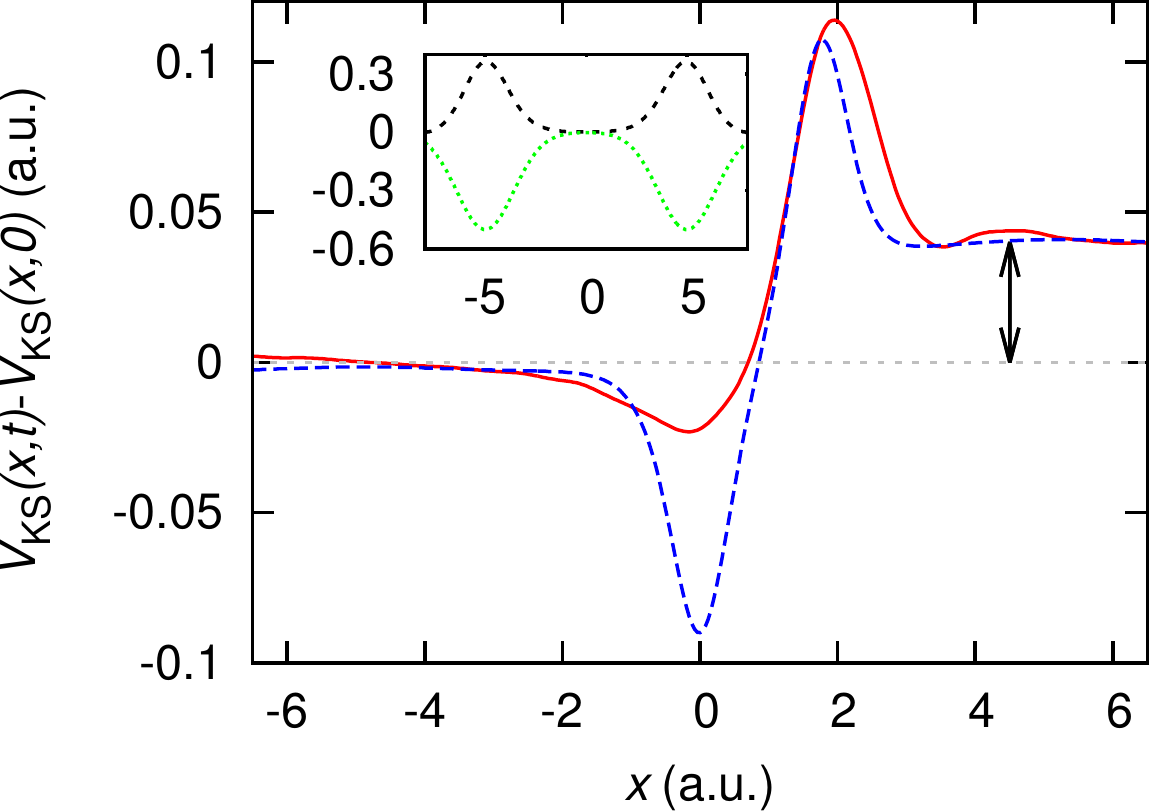}
\caption{(color online) \textit{System 4 (time-dependent double well)}. Inset: ground-state external potential (dotted green) and ground-state density (dashed black). Main panel: exact $V_{\mathrm{KS}}(x,t)-V_{\mathrm{KS}}(x,0)$, at $t=5$ a.u. (solid red), together with the corresponding quantity for the MLP (dashed blue): both replicate the potential step introduced by the time-evolution (arrow). Here $f(x,t)=0.2$.}
\label{TD}
\end{figure}

\textit{System 4 (time-dependent double well)}.---Finally we consider the extension of our functional to the time-dependent regime. Consider two electrons confined to a symmetric double well (inset of Fig.~\ref{TD}). For $t \geq 0$ an applied \textit{E}-field ($-0.1x$) forces the left electron towards the right electron. After sufficient time has elapsed ($5$ a.u.), an appreciable dynamic spatial step has formed in the corresponding KS potential between the electrons -- at the electron density minimum -- in order to replicate the effect of the Coulomb repulsion, as we showed in Ref.~\onlinecite{PhysRevB.88.241102}.
We demonstrated that there is a local dependence between the dynamic spatial steps and peaks in the velocity field ($\textbf{u} = \textbf{j}/n$), making $\textbf{u}$ the natural focus when considering appropriate functional development. We therefore extend Eq.~\ref{SOA} to include the extra terms that come about from solving the time-dependent KS equations for a single orbital, following Ref.~\onlinecite{hessler2002correlation}, giving
\begin{align}
\label{TDSOA}
V^{\mathrm{SOA}}_{\mathrm{KS}}(\textbf{r},t) = & \frac{\triangledown^2 n}{4n} -\frac{[\triangledown n]^2}{8n^2} -\int_{- \infty}^{\textbf{r}} \frac{\partial \textbf{u}}{\partial t} \cdot \mathrm{d} \textbf{r}' - \tfrac{1}{2} u^2.
\end{align}

In the MLP, $f$ in principle becomes time-dependent, but for simplicity we choose it to be constant in space and time. In order to focus on the purely time-dependent part of the MLP, we add the difference between the MLP potential at time $t$ and the MLP potential just after application of the external field at $t=0$ to the exact ground-state KS potential, and allow the KS orbitals to propagate through time, recalculating the MLP densities on each time step. 

Once again we find that a range of values for $f$, $0.1-0.3$, give accurate current and charge densities, even in the barrier region, with $f(x,t)=0.2$ most accurately replicating the potential. Figure~\ref{TD} shows $V_{\mathrm{KS}}(x,t)-V_{\mathrm{KS}}(x,0)$, at $t=5$ a.u. (the latest time studied), together with the same potential given by the time-dependent MLP for $f=0.2$. Several features are replicated by the MLP, most importantly the time-dependent xc potential step that is clearly far outside the ability of the adiabatic LDA. The details of the time-evolution of this step differ between the exact and MLP potentials, especially at earlier times. Nonetheless, the general features of the exact KS potential, as well as the time-dependent charge and current densities, are reproduced by the MLP.

In conclusion, we have shown that the single orbital  approximation (SOA) not only reproduces important non-local features of  the xc potential for systems where there is electron localization, but  also reproduces features in the KS potential in the presence of  \textit{multiple} occupied KS states. The crucial self-interaction  correction is also well described. Our mixed localization potential  (MLP), a weighted mixture of the SOA and a reference potential, extends  the good performance to regions of reduced localization. The method can  be applied self-consistently for a variety of challenging ground-state  and time-dependent situations, including the Almbladh-von Barth thought  experiment, to obtain accurate results. The MLP uses a measure of localization, $f$, for which simple approximations already give successful results for systems with various degrees of localization. Improved approximations for $f(x,t)$, building on existing DFT-based localization measures, should yield further predictive accuracy.
 
\bibliography{Bibtex}

\begin{thebibliography}{26}%
\makeatletter
\providecommand \@ifxundefined [1]{%
 \@ifx{#1\undefined}
}%
\providecommand \@ifnum [1]{%
 \ifnum #1\expandafter \@firstoftwo
 \else \expandafter \@secondoftwo
 \fi
}%
\providecommand \@ifx [1]{%
 \ifx #1\expandafter \@firstoftwo
 \else \expandafter \@secondoftwo
 \fi
}%
\providecommand \natexlab [1]{#1}%
\providecommand \enquote  [1]{``#1''}%
\providecommand \bibnamefont  [1]{#1}%
\providecommand \bibfnamefont [1]{#1}%
\providecommand \citenamefont [1]{#1}%
\providecommand \href@noop [0]{\@secondoftwo}%
\providecommand \href [0]{\begingroup \@sanitize@url \@href}%
\providecommand \@href[1]{\@@startlink{#1}\@@href}%
\providecommand \@@href[1]{\endgroup#1\@@endlink}%
\providecommand \@sanitize@url [0]{\catcode `\\12\catcode `\$12\catcode
  `\&12\catcode `\#12\catcode `\^12\catcode `\_12\catcode `\%12\relax}%
\providecommand \@@startlink[1]{}%
\providecommand \@@endlink[0]{}%
\providecommand \url  [0]{\begingroup\@sanitize@url \@url }%
\providecommand \@url [1]{\endgroup\@href {#1}{\urlprefix }}%
\providecommand \urlprefix  [0]{URL }%
\providecommand \Eprint [0]{\href }%
\providecommand \doibase [0]{http://dx.doi.org/}%
\providecommand \selectlanguage [0]{\@gobble}%
\providecommand \bibinfo  [0]{\@secondoftwo}%
\providecommand \bibfield  [0]{\@secondoftwo}%
\providecommand \translation [1]{[#1]}%
\providecommand \BibitemOpen [0]{}%
\providecommand \bibitemStop [0]{}%
\providecommand \bibitemNoStop [0]{.\EOS\space}%
\providecommand \EOS [0]{\spacefactor3000\relax}%
\providecommand \BibitemShut  [1]{\csname bibitem#1\endcsname}%
\let\auto@bib@innerbib\@empty
\bibitem [{\citenamefont {Hohenberg}\ and\ \citenamefont
  {Kohn}(1964)}]{PhysRev.136.B864}%
  \BibitemOpen
  \bibfield  {author} {\bibinfo {author} {\bibfnamefont {P.}~\bibnamefont
  {Hohenberg}}\ and\ \bibinfo {author} {\bibfnamefont {W.}~\bibnamefont
  {Kohn}},\ }\href@noop {} {\bibfield  {journal} {\bibinfo  {journal} {Phys.
  Rev.}\ }\textbf {\bibinfo {volume} {136}},\ \bibinfo {pages} {B864} (\bibinfo
  {year} {1964})}\BibitemShut {NoStop}%
\bibitem [{\citenamefont {Kohn}\ and\ \citenamefont
  {Sham}(1965)}]{PhysRev.140.A1133}%
  \BibitemOpen
  \bibfield  {author} {\bibinfo {author} {\bibfnamefont {W.}~\bibnamefont
  {Kohn}}\ and\ \bibinfo {author} {\bibfnamefont {L.~J.}\ \bibnamefont
  {Sham}},\ }\href@noop {} {\bibfield  {journal} {\bibinfo  {journal} {Phys.
  Rev.}\ }\textbf {\bibinfo {volume} {140}},\ \bibinfo {pages} {A1133}
  (\bibinfo {year} {1965})}\BibitemShut {NoStop}%
\bibitem [{\citenamefont {Perdew}\ \emph {et~al.}(1996)\citenamefont {Perdew},
  \citenamefont {Burke},\ and\ \citenamefont
  {Ernzerhof}}]{PhysRevLett.77.3865}%
  \BibitemOpen
  \bibfield  {author} {\bibinfo {author} {\bibfnamefont {J.~P.}\ \bibnamefont
  {Perdew}}, \bibinfo {author} {\bibfnamefont {K.}~\bibnamefont {Burke}}, \
  and\ \bibinfo {author} {\bibfnamefont {M.}~\bibnamefont {Ernzerhof}},\
  }\href@noop {} {\bibfield  {journal} {\bibinfo  {journal} {Phys. Rev. Lett.}\
  }\textbf {\bibinfo {volume} {77}},\ \bibinfo {pages} {3865} (\bibinfo {year}
  {1996})}\BibitemShut {NoStop}%
\bibitem [{\citenamefont {Ganduglia-Pirovano}\ \emph
  {et~al.}(2009)\citenamefont {Ganduglia-Pirovano}, \citenamefont {Da~Silva},\
  and\ \citenamefont {Sauer}}]{PhysRevLett.102.026101}%
  \BibitemOpen
  \bibfield  {author} {\bibinfo {author} {\bibfnamefont {M.~V.}\ \bibnamefont
  {Ganduglia-Pirovano}}, \bibinfo {author} {\bibfnamefont {J.~L.~F.}\
  \bibnamefont {Da~Silva}}, \ and\ \bibinfo {author} {\bibfnamefont
  {J.}~\bibnamefont {Sauer}},\ }\href@noop {} {\bibfield  {journal} {\bibinfo
  {journal} {Phys. Rev. Lett.}\ }\textbf {\bibinfo {volume} {102}},\ \bibinfo
  {pages} {026101} (\bibinfo {year} {2009})}\BibitemShut {NoStop}%
\bibitem [{\citenamefont {Perdew}\ and\ \citenamefont
  {Zunger}(1981)}]{PhysRevB.23.5048}%
  \BibitemOpen
  \bibfield  {author} {\bibinfo {author} {\bibfnamefont {J.~P.}\ \bibnamefont
  {Perdew}}\ and\ \bibinfo {author} {\bibfnamefont {A.}~\bibnamefont
  {Zunger}},\ }\href@noop {} {\bibfield  {journal} {\bibinfo  {journal} {Phys.
  Rev. B}\ }\textbf {\bibinfo {volume} {23}},\ \bibinfo {pages} {5048}
  (\bibinfo {year} {1981})}\BibitemShut {NoStop}%
\bibitem [{\citenamefont {Koentopp}\ \emph {et~al.}(2006)\citenamefont
  {Koentopp}, \citenamefont {Burke},\ and\ \citenamefont
  {Evers}}]{PhysRevB.73.121403}%
  \BibitemOpen
  \bibfield  {author} {\bibinfo {author} {\bibfnamefont {M.}~\bibnamefont
  {Koentopp}}, \bibinfo {author} {\bibfnamefont {K.}~\bibnamefont {Burke}}, \
  and\ \bibinfo {author} {\bibfnamefont {F.}~\bibnamefont {Evers}},\
  }\href@noop {} {\bibfield  {journal} {\bibinfo  {journal} {Phys. Rev. B}\
  }\textbf {\bibinfo {volume} {73}},\ \bibinfo {pages} {121403} (\bibinfo
  {year} {2006})}\BibitemShut {NoStop}%
\bibitem [{\citenamefont {Koentopp}\ \emph {et~al.}(2008)\citenamefont
  {Koentopp}, \citenamefont {Chang}, \citenamefont {Burke},\ and\ \citenamefont
  {Car}}]{DFcalc}%
  \BibitemOpen
  \bibfield  {author} {\bibinfo {author} {\bibfnamefont {M.}~\bibnamefont
  {Koentopp}}, \bibinfo {author} {\bibfnamefont {C.}~\bibnamefont {Chang}},
  \bibinfo {author} {\bibfnamefont {K.}~\bibnamefont {Burke}}, \ and\ \bibinfo
  {author} {\bibfnamefont {R.}~\bibnamefont {Car}},\ }\href@noop {} {\bibfield
  {journal} {\bibinfo  {journal} {J. Phys: Condensed Matter}\ }\textbf
  {\bibinfo {volume} {20}} (\bibinfo {year} {2008})}\BibitemShut {NoStop}%
\bibitem [{\citenamefont {Ramsden}\ and\ \citenamefont
  {Godby}(2012)}]{PhysRevLett.109.036402}%
  \BibitemOpen
  \bibfield  {author} {\bibinfo {author} {\bibfnamefont {J.~D.}\ \bibnamefont
  {Ramsden}}\ and\ \bibinfo {author} {\bibfnamefont {R.~W.}\ \bibnamefont
  {Godby}},\ }\href@noop {} {\bibfield  {journal} {\bibinfo  {journal} {Phys.
  Rev. Lett.}\ }\textbf {\bibinfo {volume} {109}},\ \bibinfo {pages} {036402}
  (\bibinfo {year} {2012})}\BibitemShut {NoStop}%
\bibitem [{\citenamefont {Elliott}\ \emph {et~al.}(2012)\citenamefont
  {Elliott}, \citenamefont {Fuks}, \citenamefont {Rubio},\ and\ \citenamefont
  {Maitra}}]{PhysRevLett.109.266404}%
  \BibitemOpen
  \bibfield  {author} {\bibinfo {author} {\bibfnamefont {P.}~\bibnamefont
  {Elliott}}, \bibinfo {author} {\bibfnamefont {J.~I.}\ \bibnamefont {Fuks}},
  \bibinfo {author} {\bibfnamefont {A.}~\bibnamefont {Rubio}}, \ and\ \bibinfo
  {author} {\bibfnamefont {N.~T.}\ \bibnamefont {Maitra}},\ }\href@noop {}
  {\bibfield  {journal} {\bibinfo  {journal} {Phys. Rev. Lett.}\ }\textbf
  {\bibinfo {volume} {109}},\ \bibinfo {pages} {266404} (\bibinfo {year}
  {2012})}\BibitemShut {NoStop}%
\bibitem [{\citenamefont {Hodgson}\ \emph {et~al.}(2013)\citenamefont
  {Hodgson}, \citenamefont {Ramsden}, \citenamefont {Chapman}, \citenamefont
  {Lillystone},\ and\ \citenamefont {Godby}}]{PhysRevB.88.241102}%
  \BibitemOpen
  \bibfield  {author} {\bibinfo {author} {\bibfnamefont {M.~J.~P.}\
  \bibnamefont {Hodgson}}, \bibinfo {author} {\bibfnamefont {J.~D.}\
  \bibnamefont {Ramsden}}, \bibinfo {author} {\bibfnamefont {J.~B.~J.}\
  \bibnamefont {Chapman}}, \bibinfo {author} {\bibfnamefont {P.}~\bibnamefont
  {Lillystone}}, \ and\ \bibinfo {author} {\bibfnamefont {R.~W.}\ \bibnamefont
  {Godby}},\ }\href@noop {} {\bibfield  {journal} {\bibinfo  {journal} {Phys.
  Rev. B}\ }\textbf {\bibinfo {volume} {88}},\ \bibinfo {pages} {241102}
  (\bibinfo {year} {2013})}\BibitemShut {NoStop}%
\bibitem [{\citenamefont {Runge}\ and\ \citenamefont
  {Gross}(1984)}]{PhysRevLett.52.997}%
  \BibitemOpen
  \bibfield  {author} {\bibinfo {author} {\bibfnamefont {E.}~\bibnamefont
  {Runge}}\ and\ \bibinfo {author} {\bibfnamefont {E.~K.~U.}\ \bibnamefont
  {Gross}},\ }\href@noop {} {\bibfield  {journal} {\bibinfo  {journal} {Phys.
  Rev. Lett.}\ }\textbf {\bibinfo {volume} {52}},\ \bibinfo {pages} {997}
  (\bibinfo {year} {1984})}\BibitemShut {NoStop}%
\bibitem [{\citenamefont {Dobson}(1991)}]{dobson1991interpretation}%
  \BibitemOpen
  \bibfield  {author} {\bibinfo {author} {\bibfnamefont {J.~F.}\ \bibnamefont
  {Dobson}},\ }\href@noop {} {\bibfield  {journal} {\bibinfo  {journal} {J.
  Chem. Phys.}\ }\textbf {\bibinfo {volume} {94}},\ \bibinfo {pages} {4328}
  (\bibinfo {year} {1991})}\BibitemShut {NoStop}%
\bibitem [{\citenamefont {Smith}\ \emph {et~al.}(1979)\citenamefont {Smith},
  \citenamefont {Jagannathan},\ and\ \citenamefont
  {Handler}}]{smith1979density}%
  \BibitemOpen
  \bibfield  {author} {\bibinfo {author} {\bibfnamefont {D.~W.}\ \bibnamefont
  {Smith}}, \bibinfo {author} {\bibfnamefont {S.}~\bibnamefont {Jagannathan}},
  \ and\ \bibinfo {author} {\bibfnamefont {G.~S.}\ \bibnamefont {Handler}},\
  }\href@noop {} {\bibfield  {journal} {\bibinfo  {journal} {International
  Journal of Quantum Chemistry}\ }\textbf {\bibinfo {volume} {16}},\ \bibinfo
  {pages} {103} (\bibinfo {year} {1979})}\BibitemShut {NoStop}%
\bibitem [{\citenamefont {Hessler}\ \emph {et~al.}(2002)\citenamefont
  {Hessler}, \citenamefont {Maitra},\ and\ \citenamefont
  {Burke}}]{hessler2002correlation}%
  \BibitemOpen
  \bibfield  {author} {\bibinfo {author} {\bibfnamefont {P.}~\bibnamefont
  {Hessler}}, \bibinfo {author} {\bibfnamefont {N.~T.}\ \bibnamefont {Maitra}},
  \ and\ \bibinfo {author} {\bibfnamefont {K.}~\bibnamefont {Burke}},\
  }\href@noop {} {\bibfield  {journal} {\bibinfo  {journal} {J. Chem. Phys.}\
  }\textbf {\bibinfo {volume} {117}},\ \bibinfo {pages} {72} (\bibinfo {year}
  {2002})}\BibitemShut {NoStop}%
\bibitem [{Note1()}]{Note1}%
  \BibitemOpen
  \bibinfo {note} {In such situations, symmetry may cause the exact KS orbitals
  to extend throughout the system. As the localized limit is approached these
  orbitals become degenerate and can be rotated within the Hilbert space onto
  effectively localized orbitals while remaining eigenfunctions of the
  Hamiltonian.}\BibitemShut {Stop}%
\bibitem [{Note2()}]{Note2}%
  \BibitemOpen
  \bibinfo {note} {We use Hartree atomic units.}\BibitemShut {Stop}%
\bibitem [{Note3()}]{Note3}%
  \BibitemOpen
  \bibinfo {note} {See Supplemental Material at [URL will be inserted by
  publisher] for specific parameters of our test systems.}\BibitemShut {Stop}%
\bibitem [{\citenamefont {Almbladh}\ and\ \citenamefont {von
  Barth}(1985)}]{almbladh1985density}%
  \BibitemOpen
  \bibfield  {author} {\bibinfo {author} {\bibfnamefont {C.~O.}\ \bibnamefont
  {Almbladh}}\ and\ \bibinfo {author} {\bibfnamefont {U.}~\bibnamefont {von
  Barth}},\ }in\ \href@noop {} {\emph {\bibinfo {booktitle} {Density Functional
  Methods in Physics}}}\ (\bibinfo  {publisher} {Springer},\ \bibinfo {year}
  {1985})\ pp.\ \bibinfo {pages} {209--231}\BibitemShut {NoStop}%
\bibitem [{Note4()}]{Note4}%
  \BibitemOpen
  \bibinfo {note} {Tests indicate that the step in the exact and SOA potentials
  always forms at the density minimum}\BibitemShut {NoStop}%
\bibitem [{\citenamefont {Gunnarsson}\ and\ \citenamefont
  {Lundqvist}(1976)}]{PhysRevB.13.4274}%
  \BibitemOpen
  \bibfield  {author} {\bibinfo {author} {\bibfnamefont {O.}~\bibnamefont
  {Gunnarsson}}\ and\ \bibinfo {author} {\bibfnamefont {B.~I.}\ \bibnamefont
  {Lundqvist}},\ }\href@noop {} {\bibfield  {journal} {\bibinfo  {journal}
  {Phys. Rev. B}\ }\textbf {\bibinfo {volume} {13}},\ \bibinfo {pages} {4274}
  (\bibinfo {year} {1976})}\BibitemShut {NoStop}%
\bibitem [{Note5()}]{Note5}%
  \BibitemOpen
  \bibinfo {note} {In this case, we can approximate Eq.~\ref {MLP+ref} as
  $V^{\protect \mathrm {MLP}}_{\protect \mathrm {KS}} = f V^{\protect \mathrm
  {SOA}}_{\protect \mathrm {KS}}+ (1-f) V_{\protect \mathrm {ext}} =
  V_{\protect \mathrm {ext}}+f V^{\protect \mathrm {SOA}}_{\protect \mathrm
  {Hxc}}$, showing that the role of $f$ is to adjust the SOA Hxc
  potential.}\BibitemShut {Stop}%
\bibitem [{Note6()}]{Note6}%
  \BibitemOpen
  \bibinfo {note} {For potentials with sufficiently high steps, the SOA
  correctly places \protect \textit {both} electrons in the lower well, even if
  each KS electron begins in a different well at the start of
  self-consistency.}\BibitemShut {Stop}%
\bibitem [{\citenamefont {Van~Gisbergen}\ \emph {et~al.}(1999)\citenamefont
  {Van~Gisbergen}, \citenamefont {Schipper}, \citenamefont {Gritsenko},
  \citenamefont {Baerends}, \citenamefont {Snijders}, \citenamefont
  {Champagne},\ and\ \citenamefont {Kirtman}}]{van1999electric}%
  \BibitemOpen
  \bibfield  {author} {\bibinfo {author} {\bibfnamefont {S.}~\bibnamefont
  {Van~Gisbergen}}, \bibinfo {author} {\bibfnamefont {P.}~\bibnamefont
  {Schipper}}, \bibinfo {author} {\bibfnamefont {O.}~\bibnamefont {Gritsenko}},
  \bibinfo {author} {\bibfnamefont {E.}~\bibnamefont {Baerends}}, \bibinfo
  {author} {\bibfnamefont {J.}~\bibnamefont {Snijders}}, \bibinfo {author}
  {\bibfnamefont {B.}~\bibnamefont {Champagne}}, \ and\ \bibinfo {author}
  {\bibfnamefont {B.}~\bibnamefont {Kirtman}},\ }\href@noop {} {\bibfield
  {journal} {\bibinfo  {journal} {Phys. Rev. Lett.}\ }\textbf {\bibinfo
  {volume} {83}},\ \bibinfo {pages} {694} (\bibinfo {year} {1999})}\BibitemShut
  {NoStop}%
\bibitem [{\citenamefont {Gonze}\ \emph {et~al.}(1995)\citenamefont {Gonze},
  \citenamefont {Ghosez},\ and\ \citenamefont {Godby}}]{PhysRevLett.74.4035}%
  \BibitemOpen
  \bibfield  {author} {\bibinfo {author} {\bibfnamefont {X.}~\bibnamefont
  {Gonze}}, \bibinfo {author} {\bibfnamefont {P.}~\bibnamefont {Ghosez}}, \
  and\ \bibinfo {author} {\bibfnamefont {R.~W.}\ \bibnamefont {Godby}},\
  }\href@noop {} {\bibfield  {journal} {\bibinfo  {journal} {Phys. Rev. Lett.}\
  }\textbf {\bibinfo {volume} {74}},\ \bibinfo {pages} {4035} (\bibinfo {year}
  {1995})}\BibitemShut {NoStop}%
\bibitem [{\citenamefont {van Faassen}\ \emph {et~al.}(2002)\citenamefont {van
  Faassen}, \citenamefont {de~Boeij}, \citenamefont {van Leeuwen},
  \citenamefont {Berger},\ and\ \citenamefont
  {Snijders}}]{PhysRevLett.88.186401}%
  \BibitemOpen
  \bibfield  {author} {\bibinfo {author} {\bibfnamefont {M.}~\bibnamefont {van
  Faassen}}, \bibinfo {author} {\bibfnamefont {P.~L.}\ \bibnamefont
  {de~Boeij}}, \bibinfo {author} {\bibfnamefont {R.}~\bibnamefont {van
  Leeuwen}}, \bibinfo {author} {\bibfnamefont {J.~A.}\ \bibnamefont {Berger}},
  \ and\ \bibinfo {author} {\bibfnamefont {J.~G.}\ \bibnamefont {Snijders}},\
  }\href@noop {} {\bibfield  {journal} {\bibinfo  {journal} {Phys. Rev. Lett.}\
  }\textbf {\bibinfo {volume} {88}},\ \bibinfo {pages} {186401} (\bibinfo
  {year} {2002})}\BibitemShut {NoStop}%
\bibitem [{\citenamefont {Van~Faassen}\ \emph {et~al.}(2003)\citenamefont
  {Van~Faassen}, \citenamefont {De~Boeij}, \citenamefont {Van~Leeuwen},
  \citenamefont {Berger},\ and\ \citenamefont {Snijders}}]{van2003application}%
  \BibitemOpen
  \bibfield  {author} {\bibinfo {author} {\bibfnamefont {M.}~\bibnamefont
  {Van~Faassen}}, \bibinfo {author} {\bibfnamefont {P.}~\bibnamefont
  {De~Boeij}}, \bibinfo {author} {\bibfnamefont {R.}~\bibnamefont
  {Van~Leeuwen}}, \bibinfo {author} {\bibfnamefont {J.}~\bibnamefont {Berger}},
  \ and\ \bibinfo {author} {\bibfnamefont {J.}~\bibnamefont {Snijders}},\
  }\href@noop {} {\bibfield  {journal} {\bibinfo  {journal} {J. Chem. Phys.}\
  }\textbf {\bibinfo {volume} {118}},\ \bibinfo {pages} {1044} (\bibinfo {year}
  {2003})}\BibitemShut {NoStop}%
\end{thebibliography}%

\end{document}